# Suppression MHD instabilities by IBW heating in HT-7 Tokamak

C.M.Qin, Y.P. Zhao, X.J. Zhang, P. Xu, Y.Yang and HT-7 team

*Institute of Plasma Physics, Chinese Academy of Sciences, Hefei 230031, PR China*

## Abstract

In HT-7 tokamak, the m= 2/1 tearing mode can be effectively suppressed by the ion bernstein wave (IBW) when the location of power deposition is near the q=2 rational surface. Off-axis electron heating and greatly increase of electron density was observed, in the meantime, the particle confinement appears to be improved with the increased of the central line averaged electron density and the drop of Da emission. Induced large ne gradients and pressures were spatially correlated with the IBW deposition profile by theoretical calculation. It is suggested that off-axis IBW heating modifies the electron pressure profile, and so the current density profile could be redistributed resulting in the suppression of the magnetohydrodynamics (MHD) instability. It provides an integrated way for making combined effects on both the stabilization of tearing modes and controlling of pressure profile.

## 1. Intoduction

Ion Bernstein waves (IBWs) have the potential to heat both ions at well-definition cyclotron harmonic resonance layers in the plasma interior and electrons via localized electron Landau damping (ELD), and also to control the plasma profiles and transport. [1] There have been a number of experiments dedicated to testing the use of IBW for fusion plasma heating,. Both ion and electron heating via linear and nonlinear processes have been observed on quite a few devices ,e.g. on JIPP-II-U[2-4], PLT[5], PBX-M[6] and Alcator-C[7]. Nevertheless, noted also the MHD instabilities were suppressed during IBW heating experiment in PBX-M, and bootstrap current induced in mid minor radius region broadened the total current profile. It has been seen the local heating resulting in modification of plasma pressure thus the current density could be redistributed [8, 9]. In addition, in HT-7 tokamak, both peaked and local steep electron pressure were realized by properly arranged the power deposition location. Suppression MHD and enhancement particle confinement during IBW heating experiments is a usual phenomenon, including low m/n tearing mode effectively suppressed by proper IBW injection [10-13]. A crucial issue for the extension of advanced tokamak scenarios to long pulse operation is to avoid these MHD instabilities [14]. This provides the possibility to apply off-axis IBW in an integrated way for making combined effects on both the stabilization of tearing modes and controlling of pressure profile.

The MHD instabilities were effectively suppressed by IBW off-axis heating in HT-7 is presented in this Letter. In Section2, the HT-7 machine and the RF heating systems are introduced. The experimental results of IBW and analysis are presented in Section 3. Finally, the conclusions are presented in Section 4.

## 2. Experiment set up

The HT-7 device is a medium-sized superconducting tokamak with a major radius of R = 1.22 m and is of a limiter configuration with two full poloidal limiters and one inner toroidal belt limiter, all of which are water-cooled and made of special doped graphite coated by SiC film with a minor radius of 27cm. In the HT-7, the ICRF heating system has a 1.5 MW continuous wave (CW) output power capability and the operation frequency range is from 18MHz to 30 MHz The ICRF system includes RF generator, transmission lines, liquid stub tuners and two antennas (FW antenna and IBW antenna). IBW antenna (see figure 1.) is a loop antenna with one end feeding and another shorted. It was installed in the horizontal mid-plane on the low field side and oriented in the toroidal direction. The radii of central conductor and Faraday shielding are 32 cm and 28.5 cm respectively. The antenna center conductor is coated by TiN, and incorporated with graphite-doped protectors and Faraday shielding to reduce impurities.

The electron temperature was measured with a sixteen-channel electron cyclotron emission (ECE) grating polychromatic (GPC) system [15]. The electron density in HT-7 was measured with a five-channel far-infrared (FIR) hydrogen cyanide (HCN) laser interferometer [16]. The exact density profile is quite difficult to be deduced from this diagnostics. Two arrays of magnetic probes which are set up on the inside of the vacuum vessel wall at different positions in the toroidal direction are used to analyses MHD activity with low m (m<5) poloidal wave numbers. Each array has 12 probes around the plasma in a circular section. From the outer-middle plane, every probe is installed at 300 with respect to the nearest neighboring probes. The frequency response of the probe is more than 100 kHz. The effective winding area is 1430 cm2.

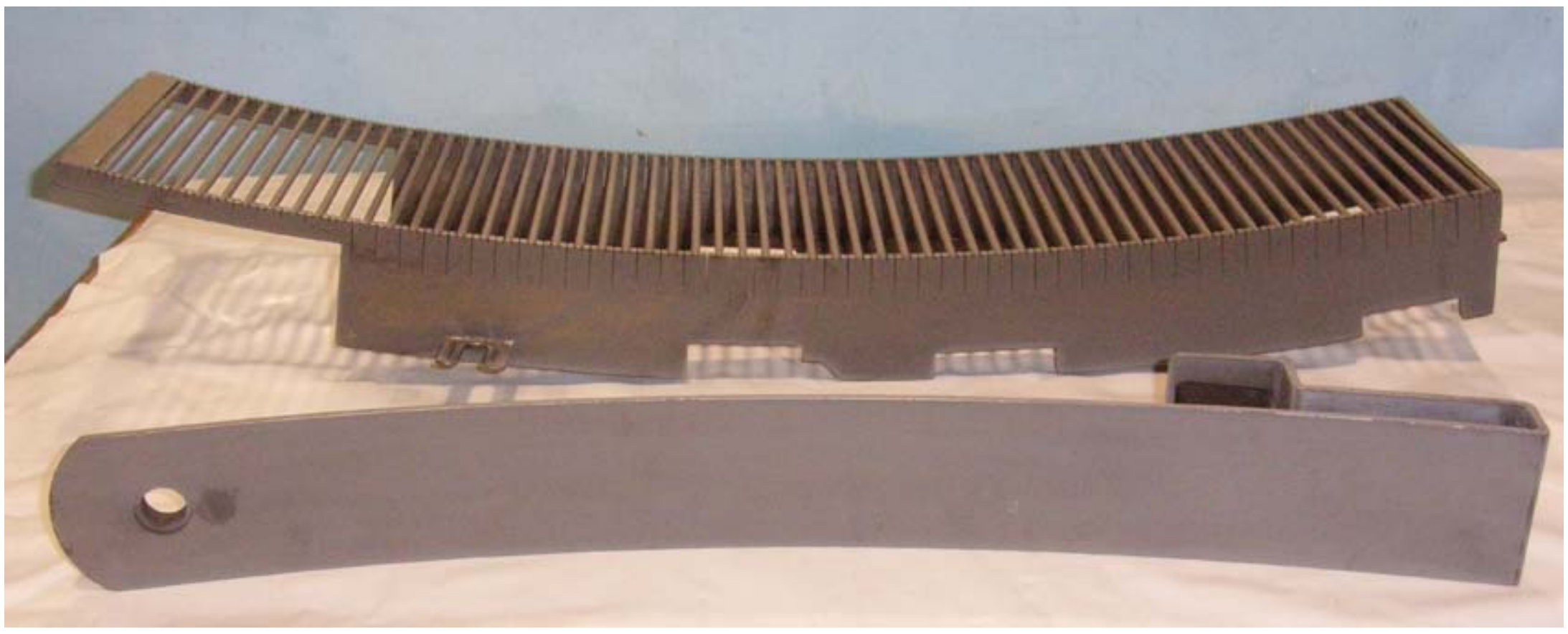

Figure1. IBW antenna on HT-7 tokamak, installation inside the vessel of low field side

## 3. Experimental results

In HT-7, MHD instabilities occur frequently in the some OH operation regime with plasma current of 100-120kA, electron density of $0.9\text{-}1.1\cdot 10^{13}\text{cm}^{-3}$ and toroidal magnetic field of 1.7-2.0Tesla. During the plasma current platform, most of them are dangerous m/n=2/1 resistive tearing mode, which is driven by the plasma current density gradient. It interacts with the vacuum vessel and often intervenes to completely lock the mode, which leads to disruption in high-power-heated plasmas. Suppression or avoidance of the m/n = 2/1 resistive tearing mode through current density redistribution was a general way in HT-7.

In HT-7, the investigation on 2/1 resistive tearing mode suppression by IBW has bee performed. The experiments were conducted in deuterium target plasmas with plasma current Ip = 120 kA, and central line-averaged electron center density.ne = $1.1\times 10^{19}$ $\text{m}^{-3}$. The ratio of H/(H+D) is about 20%. IBWs with frequency of 30 MHz are used in the experiments. For this frequency, the peak of the parallel refractive index spectrum, $n_{//}$ of IBW is around 8, and the power component in the low $n_{//}$ portion is weak [13]. Such a parallel spectrum is favorable for electron heating via ELD [17].

Scanning of the toroidal magnetic field from 1.7-2.0Tesla, the resonance location of 30MHz

IBW was varied across the plasma minor radius, from plasma edge to plasma center. It was found when the location of the $2\Omega_D$ or $\Omega_H$ ion cyclotron resonant layer is closer to the q=2 rational surface, the 2/1 tearing modes can be effectively suppressed by IBW off-axis heating .

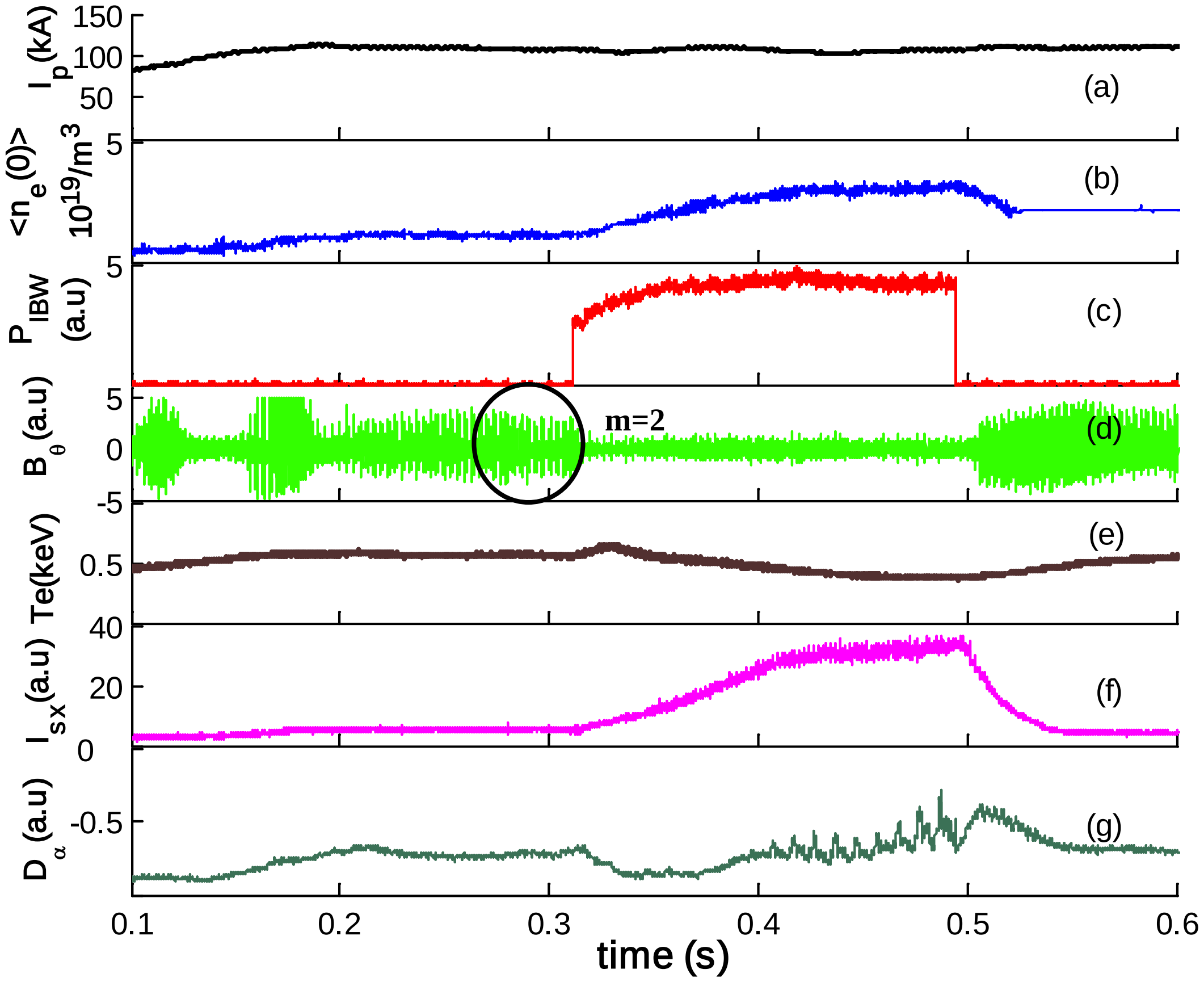

Figure 2. A typical MHD instabilities suppressed by IBW (shot No 98061) in the HT-7. (a) Plasma current; (b) central line average electron density ; (c) the IBW power P=160kW; (d) ECE intensity from channel; (e) Mirnov signals; (f ) the intensity of soft x-ray at center; (g )Da emission;

The typical discharge of MHD stabilization with 1.75 Tesla toroidal field ($B_T$) and 30 MHz IBW heating is shown in figure 2. The power of IBW was about 140 kW at flat top. The amplitude of Mirnov oscillation greatly decreased throughout the IBW duration. Obviously, the MHD instabilities was effectively suppressed by IBWs. Magnetic analysis by singular value decomposition method (SVD) [18-19] for the discharge in Fig.2 shows that the MHD activity consists predominantly of m= 2 modes with a rotation frequency of about 7.8 kHz, (see Fig. 3). After knowing the current density profile, $J \sim (1-X^2)^{\alpha}$ ,where $X = r/a$ . The q-profile can be calculated by using the formula: $q(r) = q(a)^{X^2} / [1-(1-X^2)^{\alpha+1}]$, For the shot no. 98061, $q(a) = 3.0$, and $\alpha = 3$, the corresponding q(r) profile is shown as Fig.4. In the range of the MHD activity, *m*= 2 modes, the resonant layer of *q* is located at around 0.55a. In the case of IBW with 30MHz frequency and 1.75 Tesla magnetic field strength, the $2\Omega_D$ or $\Omega_H$ ion cyclotron resonant layer at a radius of 13 cm (~0.5a) is very close to the q=2 rational surface. The time traces of magnetic fluctuation and different IBW power injection with two adjacent shots as shown in Fig.9. It is seen the effect of tear mode suppression is better than lower IBW power injection. It is found the power threshold of tearing mode stabilization could be around 100kW with the scanning of the IBW power.

The central line-averaged electron density (Fig. 2(e)) was increased 2.5 factor during IBW heating. The electron temperature of the plasma centre measured by electron cyclotron emission (ECE) slightly increased in the beginning 40 ms after injection of IBW. (Fig. 2(e)), The heat exchange time between electron and deuterium ion was about 100 ms. The electron should be directly heated by IBW. It can be seen in figure 5 the electron at plasma center and at r~13cm were heated by IBW and the profile of electron temperature become more peaked compared to that at OH heating. But the electron temperature of whole plasma dropped very much when electron density reached the flat top. The central soft x-ray signal (Fig. 2(f)) was strongly increased during IBW heating. The particle confinement appears to be improved indicating by a density increase and a drop of the intensity of the Da radiation (Fig. 2(g)). The polodial shear flow induced by IBW maybe the reason for improved particle confinement [20-21].

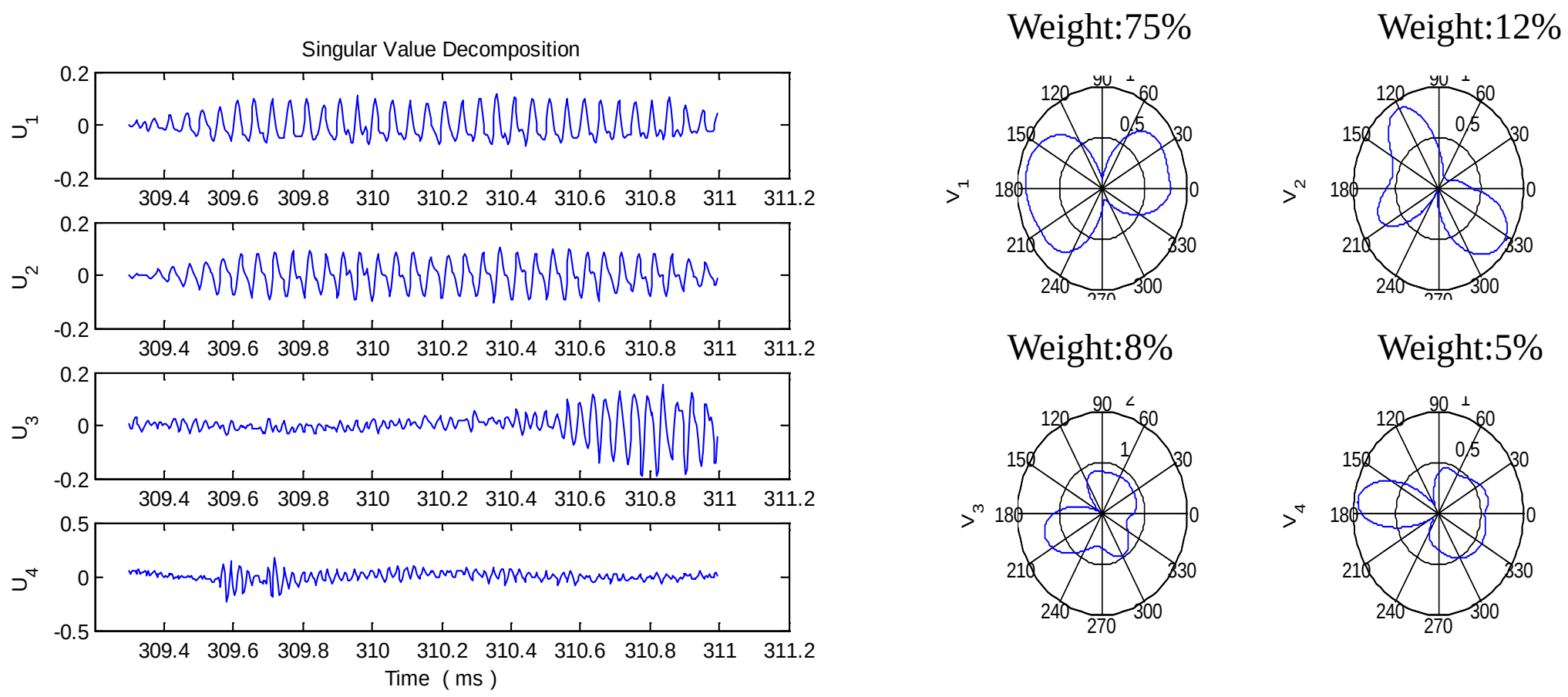

(a) (b)

Figure.3 Results of the SVD analysis from magnetic signals. (a) the first order temporal eigenvector of the mode, (b) the spatial eigenvector corresponding to the temporal eigenvector in Fig. 3(a). The main mode number is m=2.

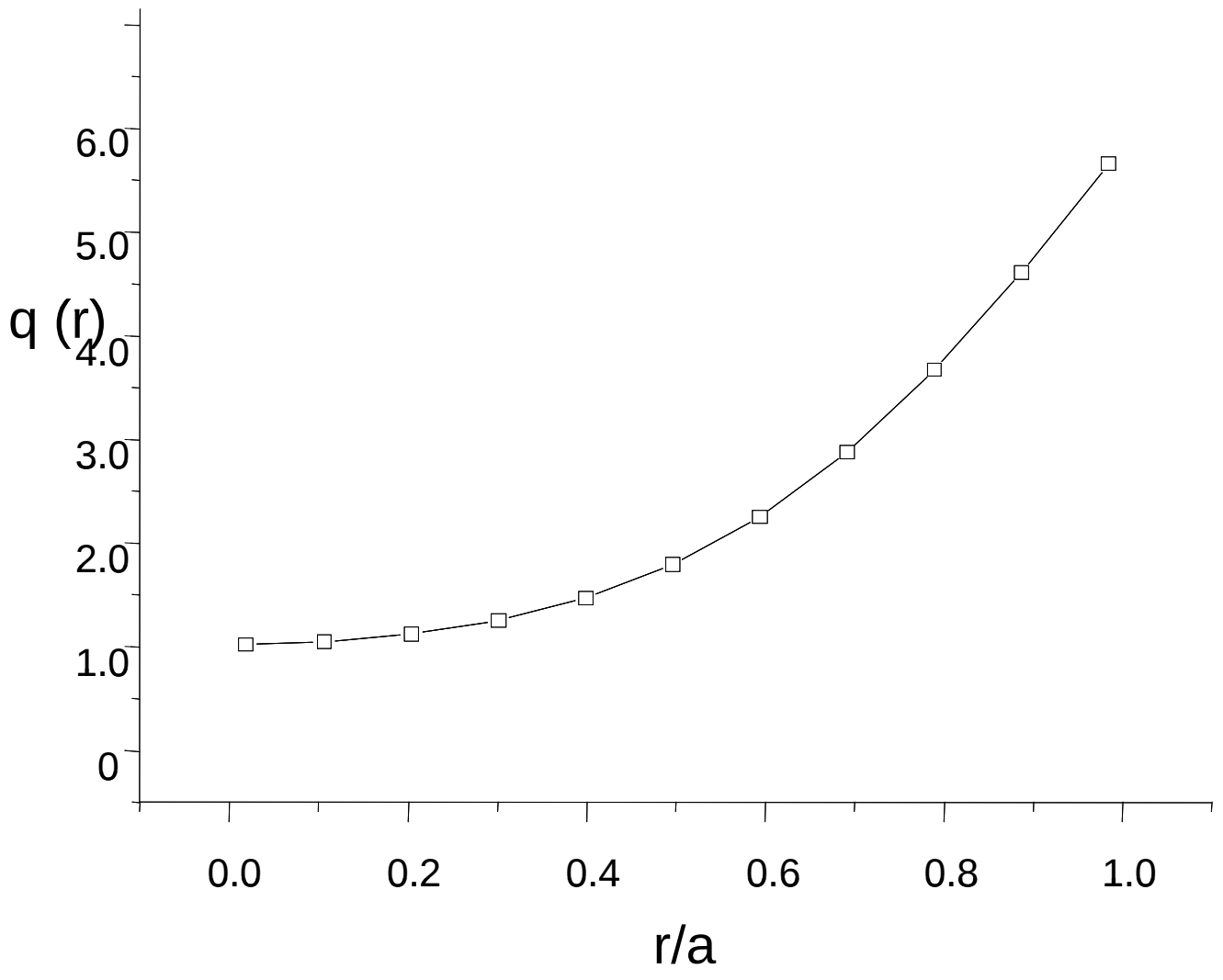

Figure.4 Calculated q(r) profile for discharge shot No. 98061.

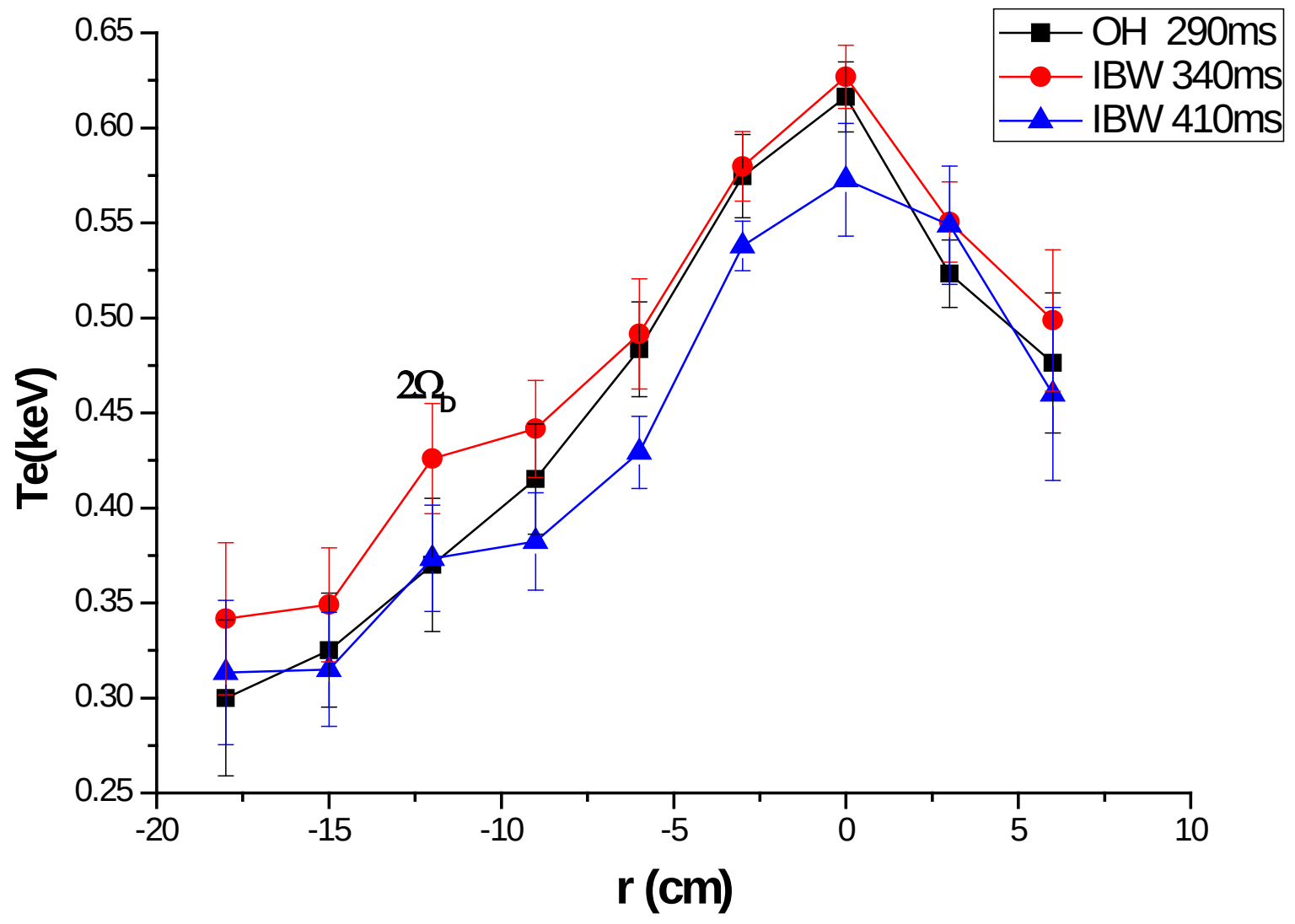

Figure 5. electron temperature profile in Ohmic (290ms) and IBW heating (340ms and 410ms)phases. The electron temperature profiles were measured by SX-PHA.

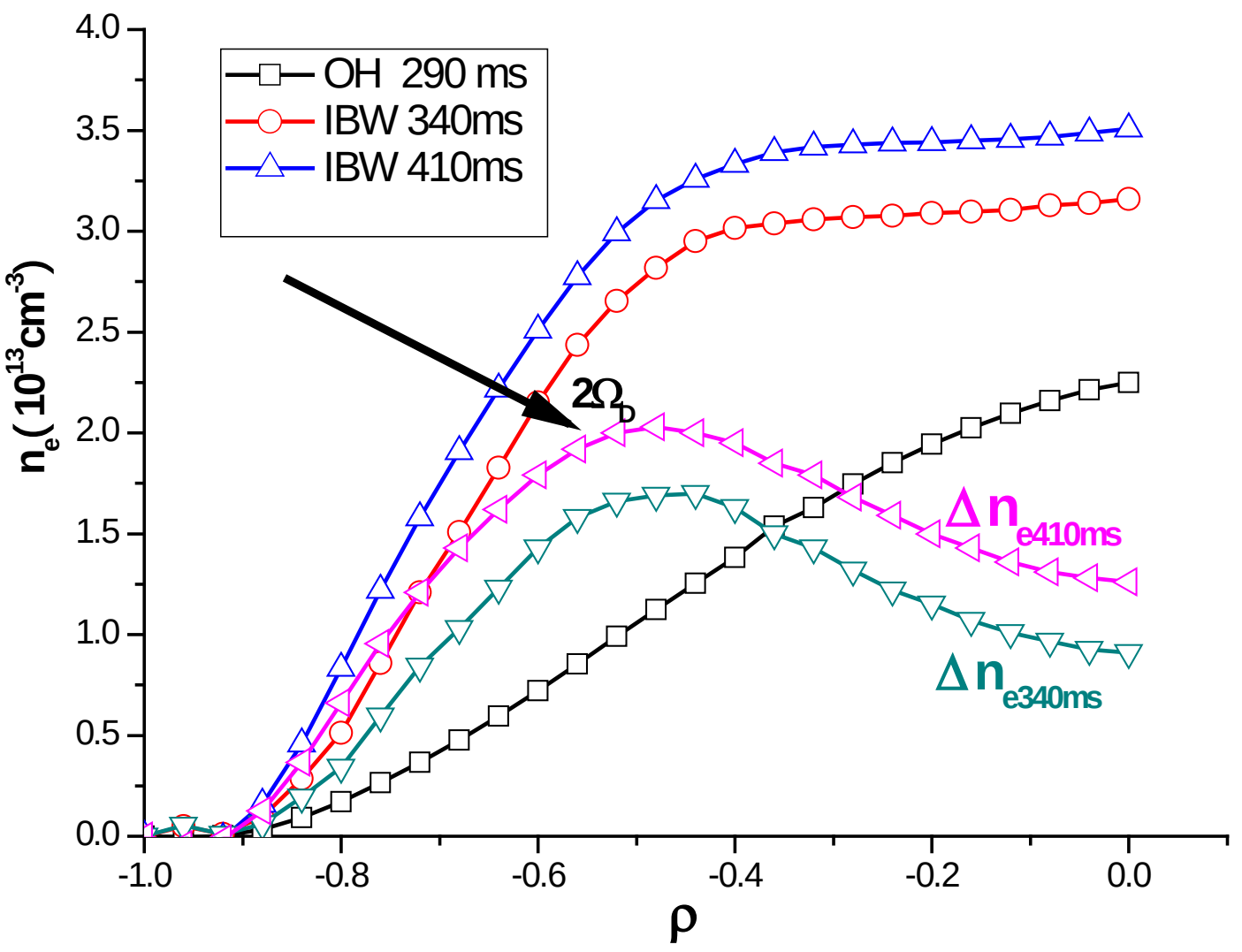

Figure 6. The Abel inverted electron density profiles in Ohmic (290ms) and IBW heating (340ms and 410ms) phases, Δne is the increment of electron density during IBW phasing comparing to OH phasing.

The electron density profile strongly increased and peaked during the IBW heating as shown in Fig. 5 at 0.29 s for ohmic heated phase and 0.34 s,0.41s for IBW heated phase during the electron density plateau, where ρ is the normalized minor radius. The electron density increases by a factor up to 2.5 and the density profile broadens during IBW heating. Similar peaking of the density profile has been also observed on JIPPT-II-U [22] and PBX-M [8-9] IBW experiments. It is noted that the maximum electron density increment at the mid minor radius, ρ ~ 0.5a (around r = 13.5cm) with IBW heating. Which corresponds to the $2\Omega_D$ or $\Omega_H$ ion cyclotron resonant layer at a radius of 13 cm. Due to the slightly increase of electron temperature in the same region, the maximum pressure increment also occurred at the power deposition region. This area is well coincide with the IBW power deposition profile calculated by a ray tracing (see fig.7). Theoretical calculations predicated that the IBW power mainly deposits on electrons with most power absorbed within $\rho \sim 0.5a$, localized at an off-axis radius. The wave power is absorbed by electrons due to ELD which is strongly related to the parallel refractive index of the launched wave power spectrum before the IBW reaches the ion cyclotron resonance layer. But ELD at the maximum $n_{//}$ was relative weak because the electron temperature was low in the outer half minor radius region. Since more power was available for ELD before reaching the resonant layer, a larger increment of localized electron temperature resulted [23]. This feature can be used externally to control the plasma electron pressure profiles, if the plasma target and the launched IBW parallel spectrum are properly arranged. In HT-7 and PBX-M, the controlling of pressure profile is well realized with IBWs [24-27].

The experimental results show that the MHD activity was indeed suppressed by IBW off-axis heating. Due to the power absorption is localized and peaked density on the off-axis power deposition region. It can modify the pressure profile at the region, as result in a larger bootstrap current fraction, thus current density could be redistributed. Indeed, similar results of modification current density profile IBW and MHD stabilities were found in the PBX-M tokamak [8, 28]. The ability for IBW to change the density profile appears to be particularly attractive for controlling the bootstrap current profile for advanced tokamaks[26,27].

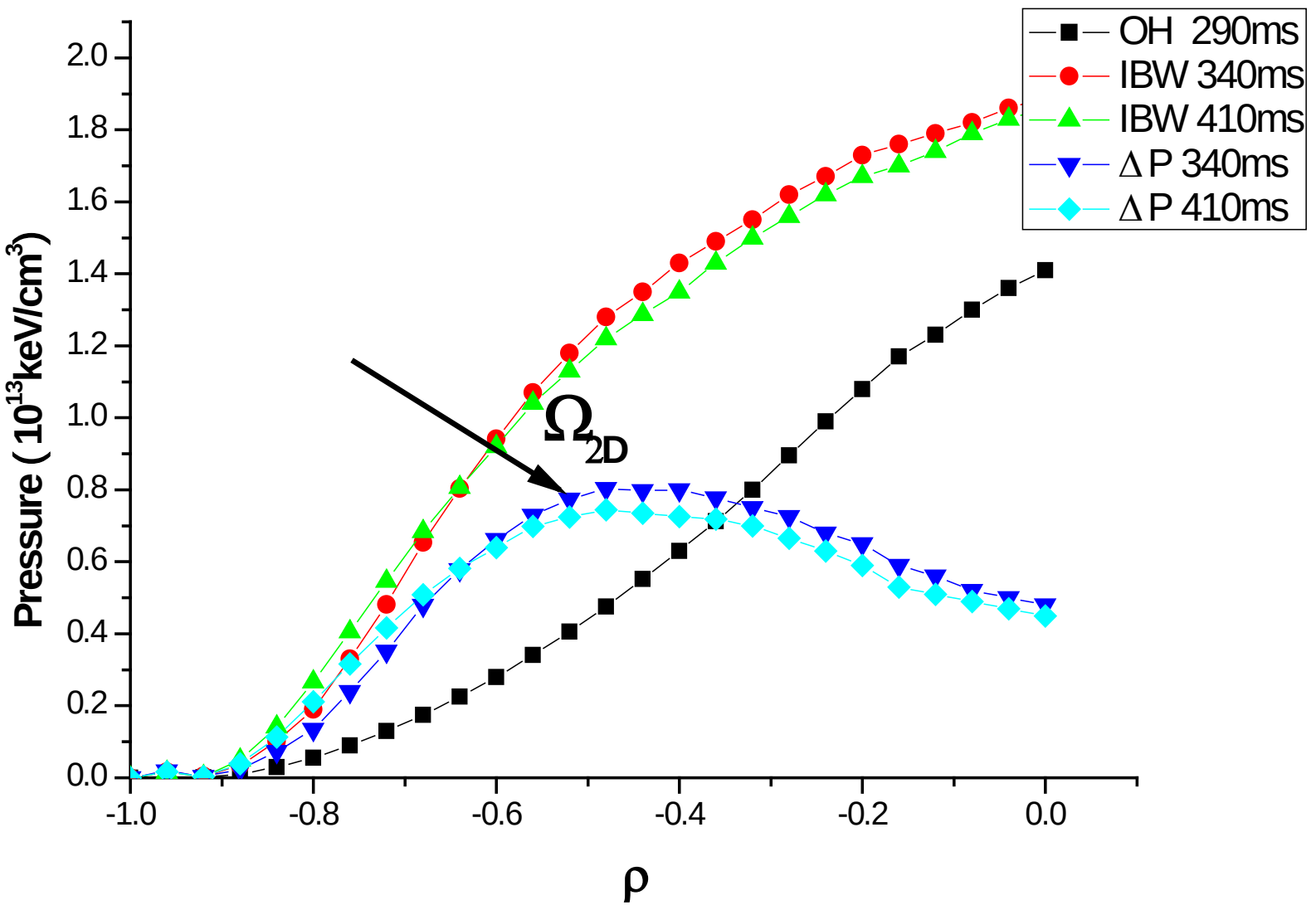

Figure 7. Pressure profiles in Ohmic target plasma and IBW heating phases. ΔP is the increment of pressure during IBW phase.

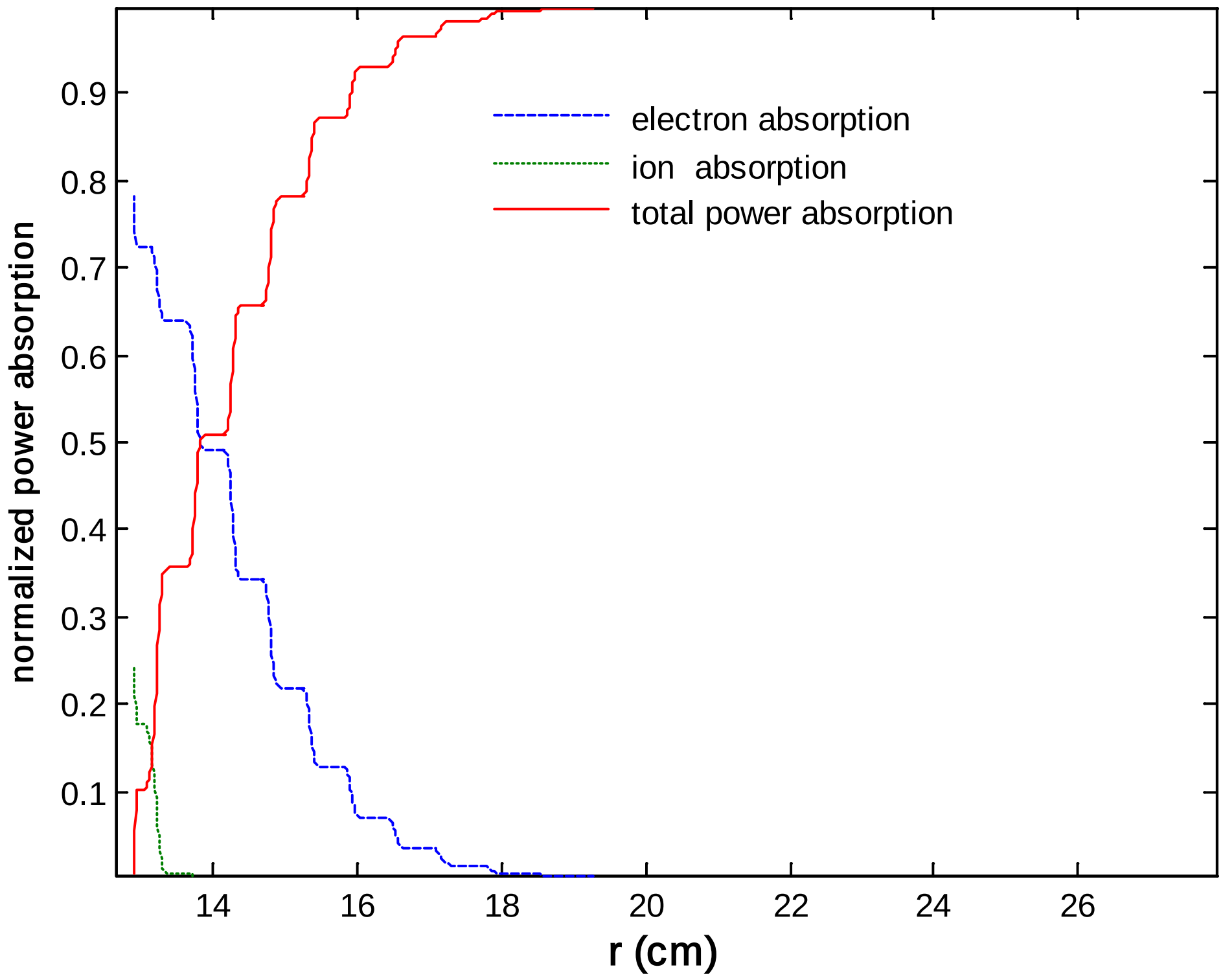

Figure8. The simulated radial power density deposited profile of IBW using a ray tracing code for HT-7. While the power is deposited off-axis in the discharge, with the parameters: ne(0)= 1.0 x $10^{19}m^{-3}$,nedge= 0.5 x $10^{19}m^{-3}$ , Te(0)= 700eV, Ti(0)= 700 eV, Te(e)= Ti(e)= 40 eV, B0=1.78 Tesla, f=30 MHz

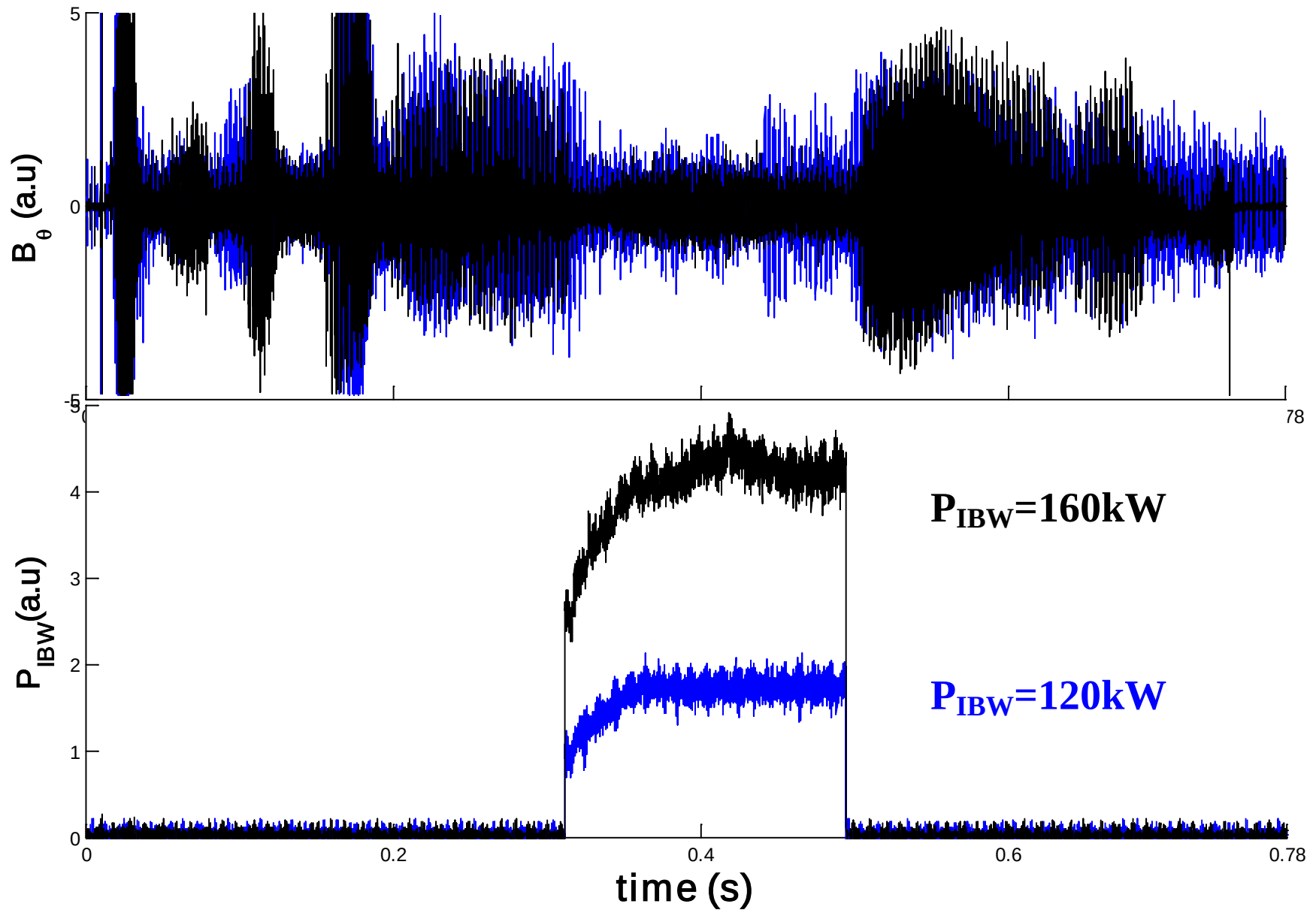

Figure 9. Time traces of magnetic fluctuation and different IBW power injection with two shots: no. 98060 and no. 98061.

## 4. Discussion and Summary

A preliminary analysis shows that electron heating and electron density increment were spatially localized and in agreement the IBW power deposition profile calculated by a ray tracing code. The electron pressure profile can be externally controlled by IBW in a well-defined local region. Successfully m=2 tearing mode suppression was achieved when the IC resonance was located in the region near the q=2 rational surface. The power threshold of suppression of MHD instablization could be around 100kw. The contribution to stabilization mainly comes from the IBW power deposition in the vicinity of the q = 2 rational surface, modification the pressure profile, and current density could be redistributed, such as j ( r ) profile flattening thus effect MHD behaviors. It is a remaining issue, which needs to be further investigated in HT-7. This feature provides the possibility to apply off-axis IBWH in an integrated way for making combined effects on both the stabilization of tearing modes and controlling of pressure profile.

**Acknowledgements**

This work was supported by the National Natural Science Foundation of China under Grant No. 10675125